\documentclass[twocolumn,prc,aps,showpacs]{revtex4}
\usepackage{graphicx}

 \def\gsim{\mathrel{\rlap{\lower4pt\hbox{\hskip1pt$\sim$}}
 \raise1pt\hbox{$>$}}}

 \newcommand\la{\langle}
 \newcommand\ra{\rangle}
 \newcommand\beq{\begin{equation}}
 
 \newcommand\eeq{\end{equation}}
 \newcommand\beqn{\begin{eqnarray}}
 \newcommand\eeqn{\end{eqnarray}}
\def\mb{\,\mbox{mb}}

\def\GeV{\,\mbox{GeV}}

\def\lsim{\mathrel{\rlap{\lower4pt\hbox{\hskip1pt$\sim$}}
    \raise1pt\hbox{$<$}}}         
\def\gsim{\mathrel{\rlap{\lower4pt\hbox{\hskip1pt$\sim$}}
    \raise1pt\hbox{$>$}}}         

\def\mb{\,\mbox{mb}}

\def\GeV{\,\mbox{GeV}}

\def\s0{\sigma_0(s)}

\def\beq{\begin{equation}}
\def\eeq{\end{equation}}

\def\beqy{\begin{eqnarray}}
\def\eeqy{\end{eqnarray}}

\newcommand{\ber}{\begin{displaymath}}
\newcommand{\eer}{\end{displaymath}}
\newcommand{\bey}{\begin{eqnarray}}
\newcommand{\eey}{\end{eqnarray}}

\def\beq{\begin{equation}}
\def\eeq{\end{equation}}

\def\beqy{\begin{eqnarray}}
\def\eeqy{\end{eqnarray}}

\begin{document}

\title{\bf Penetrating intrinsic charm: evidence in data}

\vspace{1cm}

\author{B. Z. Kopeliovich}
\author{I. K. Potashnikova}
\author{Iv\'an Schmidt}
\affiliation{Departamento de F\'{\i}sica, Centro de Estudios
Subat\'omicos, Universidad T\'ecnica Federico Santa Mar\'{\i}a,
\\and\\
Centro Cient\'ifico-Tecnol\'ogico de Valpara\'iso,\\
Casilla 110-V, Valpara\'iso, Chile}

\begin{abstract}
\noindent
Nuclei are transparent for a heavy intrinsic charm (IC) component of the beam hadrons, what leads to an enhanced nuclear dependence of open charm production at large Feynman $x_F$. Indeed, such an effect was observed recently in the SELEX experiment \cite{selex}. Our calculations reproduce well the data, providing strong support for the presence of IC in hadrons in amount less than $1\%$. Moreover, we performed an analysis of nuclear effects in $J/\Psi$ production and found a similar, although weaker effect.

\end{abstract}



\maketitle
\vspace{1cm}

All data available so far for particle production at forward
rapidities in high-energy hadron-nucleus collisions exhibit a
similar trend for enhanced nuclear suppression at larger Feynman
$x_F$ (e.g. see the collection of data in \cite{forward}). However,
the SELEX experiment has released recently data for charmed hadron
production in hadron-nucleus collisions  \cite{selex}, which
demonstrate an opposite behavior: the nuclear ratio is rising,
rather than  falling, at large $x_F$. Here we interpret this effect
as a manifestation of the projectile intrinsic charm (IC) component,
which does not attenuate propagating through the nucleus. So it is
filtered out and relatively enhanced by the nucleus. Such a signal
of IC can also be observed in data for nuclear dependence of
$J/\Psi$ production.

{\it Charmed hadron production.} First of all, one should understand
why produced hadrons are suppressed by nuclei, as one can see in the
SELEX data \cite{selex} for charmed hadrons, as well as in data for
other hadronic species \cite{forward}. According to the
Mueller-Kancheli theorem \cite{mueller, kancheli}, inclusive cross
sections of particle production, $hA\to h'X$, should have no
shadowing. This conclusion is based on the
Abramovsky-Gribov-Kancheli (AGK) cutting rules \cite{agk} with an
extra factor which is equal to the number of cut Pomerons.   In the
particular case of hadron-nucleus collisions one can expand the
inclusive cross section over the number of collisions $n_{coll}$,
with the Glauber weights and with an extra factor, $n_{coll}$,
compared to the same expansion for the total cross section. This is
because the final hadron $h'$ can emerge from any of the multiple
inelastic collisions. Then, instead of the $A^{2/3}$ characteristic
behavior of the total cross section, one obtains a linear, $\propto
A$, rise of the inclusive cross section.

This prediction has been indeed confirmed in hard reactions having
small cross sections, like the Drell-Yan process, production of
hadrons and direct photons with high transverse momentum, etc.
However, in soft precesses one should expect the linear $A$
dependence to be broken due to saturation of the unitarity bound.
Namely, hadron production is enhanced by multiple interactions only
if the cross section is far from the unitarity bound. Otherwise, the
production processes on different nucleons shadow each other and the
cross section remains essentially unchanged by multiple
interactions, an effect known as Landau-Pomeranchuk principle (see
in \cite{broad}). Only the transverse momentum distribution is
modified, the called color glass condensate \cite{mv}. It is also
related to coherence  in particle production from multiple sources:
when the phase space of produced particles is densely packed up to
some transverse momentum  $Q_s$, the inclusive cross section
saturates with number of collisions, i.e., the inclusive cross
section is independent of number of collisions. Then, one should
suppress the extra factor $n_{coll}$ in the Glauber expansion, and
eventually arrive at a shadowed inclusive cross section $\propto
A^{2/3}$.

Moreover, at large Feynman $x_F\to1$ the nuclear suppression is
unavoidably getting stronger due to the dissipation of energy in
multiple interactions in the nucleus, and the restrictions imposed
by energy conservation. Eventually, near the kinematic limit the
suppression reaches the maximal possible magnitude corresponding to
the $A^{1/3}$ dependence of the cross section. One can interpret
this mechanism as excitation of higher Fock components in the beam
hadron by multiple interactions in the nucleus. The more
constituents a Fock state has, the steeper is the fall of the parton
distribution function at $x\to1$. Thus, each of the multiple
collisions, but the first one, supplies an  additional suppression
factor $S(x_F)$ \cite{forward}, and summing up the modified Glauber
expansion over number of collisions, one gets \beqn A_{eff}(x_F)&=&
\frac{1}{S(x_F)\sigma_{abs}} \int d^2b\,e^{-\sigma_{abs}T_A(b)}
\nonumber\\ &\times& \left[e^{S(x_F)\sigma_{abs}\,T_A(b)} -1\right].
\label{100}
 \eeqn
Here the effective atomic number $A_{eff}(x_F)=\sigma^A(x_F)/\sigma^N(x_F)$ is the ratio of the inclusive cross sections. The nuclear thickness function $T_A(b)=\int_{-\infty}^{\infty}dz\,\rho_A(b,z)$ is the integral of the nuclear density $\rho_A(b,z)$ along the trajectory at impact parameter $b$.

The suppression factor $S(x_F)$ should vary from 1 at $x_F=0$ (see
above) down to vanishing as $1-x_F$ at $x_F\to1$
\cite{brodsky,forward}. Here we rely on the simplest form
$S(x_F)=1-x_F$, which describes well a bulk of data for production
of light hadrons \cite{forward}. The effective absorption cross
section $\sigma_{abs}$, which controls the number of collisions, in
the Glauber approximation would be just the inelastic hadron-nucleon
cross section. However, the Gribov inelastic corrections
\cite{gribov} make the nuclear medium more transparent and lead to a
considerable reduction of $\sigma_{abs}$ compared to
$\sigma^{hN}_{in}$ \cite{mine}. Similar to Ref.~\cite{forward}, we
evaluated the effective absorption cross section by comparing the
Glauber exponential attenuation with one calculated in the dipole
approach (see Eq.~(58) in \cite{kps1}). For this evaluation we use
the mean nuclear thickness for each nucleus. For $\Sigma^-$ hyperon
and pion we use the results of the SELEX experiment
\cite{selex-tot}, $\sigma^{hp}_{tot}/\sigma^{pp}_{tot}=0.9$ and
$0.635$ for $h=\Sigma^-$ and pion, respectively. This procedure was
applied to copper and carbon separately and as function of energy,
in accordance with data in Ref.~\cite{selex}. For example, for
copper at $p_{lab}=300\GeV$ we obtained $\sigma^{p}_{abs}=20.4\mb$,
$\sigma^\pi_{abs}=13.1\mb$ and $\sigma^{\Sigma^-}_{abs}=18.8\mb$ .
Notice that with this value of $\sigma^{p}_{abs}$ Eq.~(\ref{100})
describes well the $x_F$ dependence of nuclear suppression observed
for many hadronic species produced in $pA$ collisions
\cite{forward}.

\begin{figure}[htb]
\includegraphics[width=6cm]{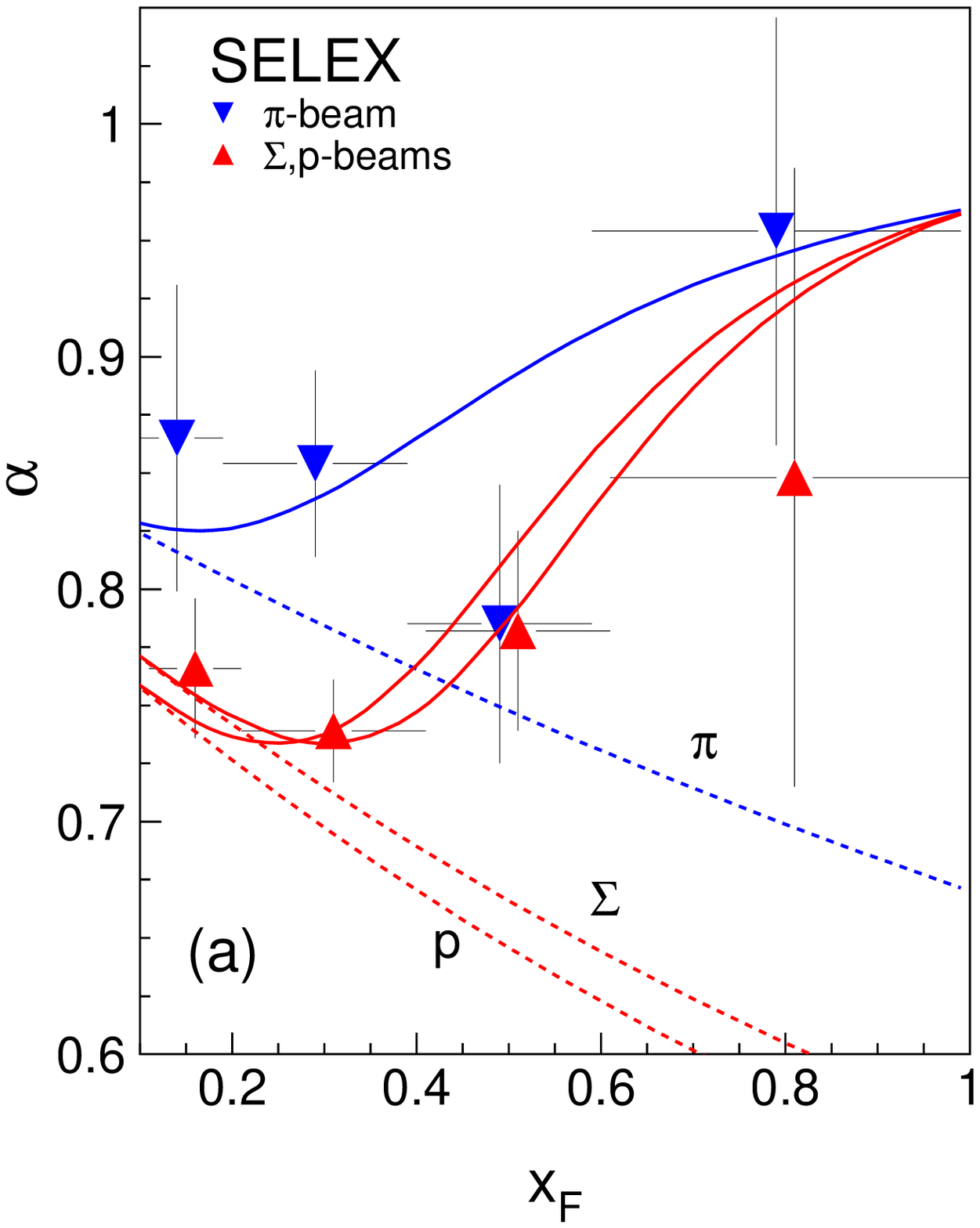}\\[4mm]
\includegraphics[width=6cm]{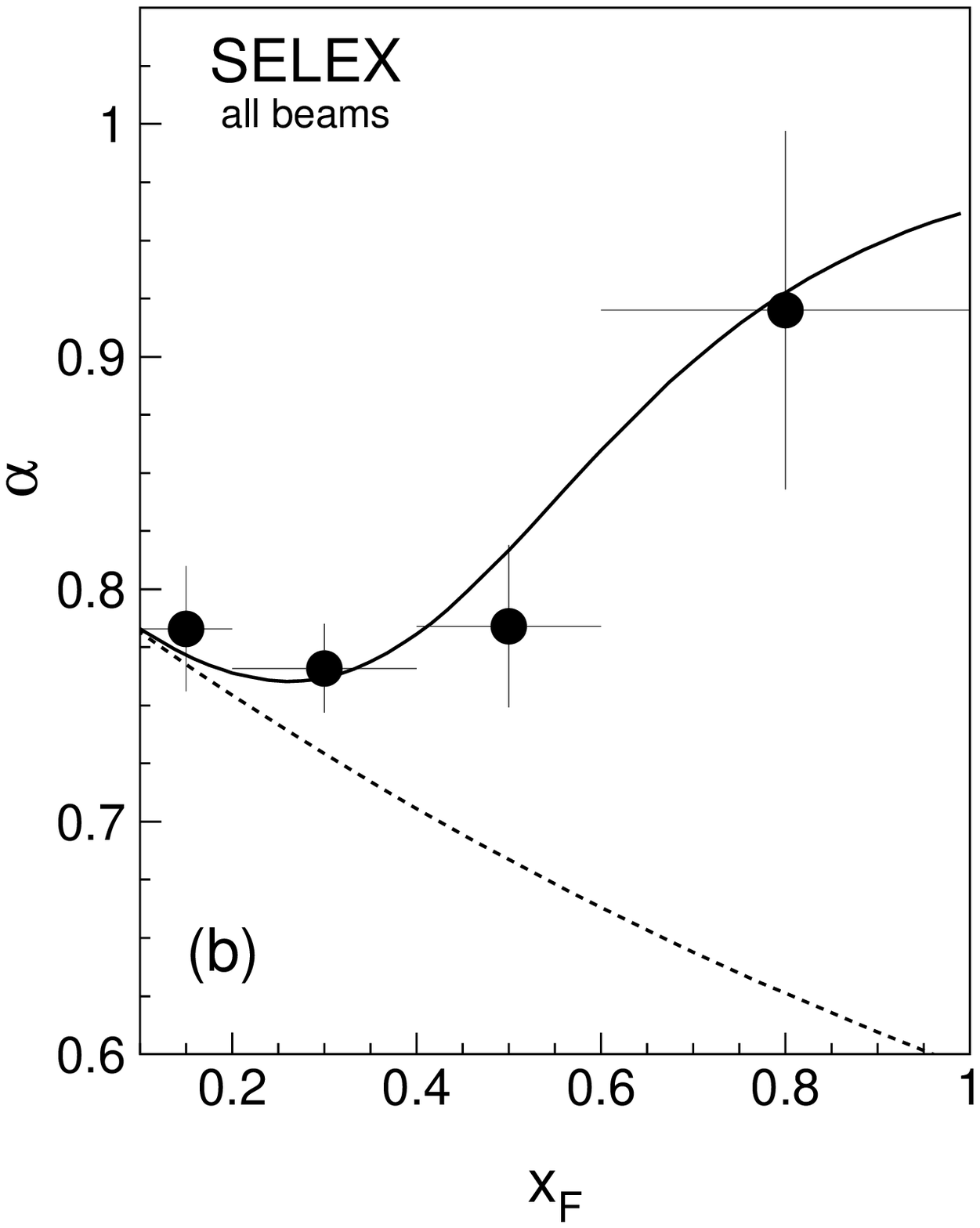}
\caption{\label{alpha1} {\bf (a)} The exponent $\alpha(x_F)$ characterizing the $A$-dependence of the cross section
as function of $x_F$. The dashed curves show $\alpha(x_F)$ for the conventional (no IC) mechanism of light quark fragmentation calculated with Eq.~(\ref{100}). Solid curves show the effect of inclusion of IC as is given by Eq.~(\ref{190}). The curves from bottom to top correspond to proton, $\Sigma^-$ and pion beams, respectively. {\bf (b)} Same as above, but for the combined statistics for all beams.
}
 \end{figure}

In Fig.~\ref{alpha1} our calculations for the exponent $\alpha$ are
shown by dashed curves in comparison with the SELEX data.
While at small $x_F$ our results agree with the data,
Eq.~(\ref{100}) underestimates the value of $\alpha$ measured at
large $x_F$. This anomaly might signal a missed IC contribution,
since IC in  light hadrons carries the main fraction of its momentum
\cite{ic1,ic2,ic3,ic4} and becomes a new source of charmed hadrons
at large $x_F$. Since the mechanism of fragmentation of a light
projectile quark is nuclear suppressed at large $x_F$, the
fragmentation of a fast charm quark originating from the IC becomes
more important.

A peculiar feature of heavy quarks is their weak attenuation in a medium. The mean momentum fraction of an intrinsic heavy component remains large even in higher Fock components, provided that the heavy quark mass squared is considerably larger than the square of the hadron mass. For the same reason the medium induced perturbative energy loss is much less for heavy than for light quarks \cite{dokshitz}.
Thus, multiple nuclear interactions, which excite higher Fock states in the projectile hadron, do not affect much the IC momentum distribution and do not suppress the originated from IC charmed hadrons at large $x_F$. This is the key observation, which makes the IC a plausible mechanism responsible for the anomalous nuclear dependence observed in the SELEX experiment.

\begin{figure}[htb]
 \includegraphics[width=8cm]{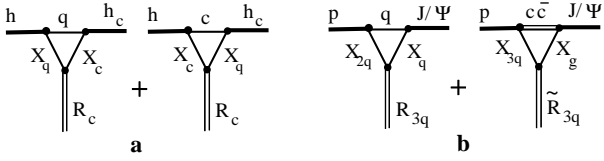}
\caption{\label{vertex} {\bf a:} Regge vertex $h$-$h_c$-$R_c$ for two mechanism of $h_c$ production. $X_c$ and $X_q$ denote quark assembles with or without open charm, respectively.
{\bf b:} The vertex $p$-$J/\Psi$-Reggeon for two mechanisms of $J/\Psi$ production, $q\bar q$ annihilation (left), and either the perturbative glue-glue fusion,
or the IC contribution (right). $X_{3q}$ and $X_{2q}$ are the three or two quark partonic systems, respectively; $X_g$ is a pure gluonic system.
 }
 \end{figure}

To proceed further with the interplay of the two mechanisms of charmed hadron production, notice that
the limiting behavior of the cross section at $x_F\to1$
is independent of presence or absence of IC, since it is controlled by the triple Regge formalism,
\beq
\left.\frac{d\sigma(hp\to h_cX)}{dx_F\,dt}\right|_{x_F\to1}\propto
(1-x_F)^{1-2\alpha_{R_c}(t)},
\label{150}
\eeq
where $t$ is the 4-momentum transfer squared, and
$\alpha_{R_c}(t)$ is a charmed Regge trajectory which assignment depends on the quantum numbers of $h$ and $h_c$. Only the structure the $h$-$h_c$-$R_c$ vertex depends on the production mechanisms under discussion, as is illustrated in Fig.~\ref{vertex}a, but the Reggeon remains the same.

The $t$-integrated cross sections Eq.~(\ref{150}) at $x_F\to1$ has the form
$d\sigma/dx_F\propto(1-x_F)^n$, which is the universal behavior for both mechanisms.
We denote the relative IC contribution (Fig.~\ref{vertex}a right) by $\delta$, correspondingly the rest
(Fig.~\ref{vertex}a left) is $1-\delta$. While these two mechanisms have the same $x_F$ dependence at $x_F\to1$, the IC contribution declines from this behavior at smaller $x_F$. In particular, the IC contribution is expected to have a maximum at $x_F\approx \la z_{IC}\ra/2$, where $\la z_{IC}\ra= 0.75$ is the position of the peak in the IC distribution function \cite{higgs}.
Thus, we assume the $x_F$-dependence of $h_c$ production  to be shaped as
\beq
\left.\frac{d\sigma(hp\to h_cX)}{dx_F}\right|_{IC}=
\delta\,x_F^m\,\frac{d\sigma(hp\to h_cX)}{dx_F},
\label{180}
\eeq
where $m=\la z_{IC}\ra\,n/(2-\la z_{IC}\ra) = {3\over5}\,n$.

The exponent $n$ controlling the asymptotic behavior of charmed hadron production at $x_F\to1$ can be expressed via the intercepts of the relevant Regge trajectories. For instance, $n(\pi^-\to D^0,D^-)=1-2\alpha_{D^*}(0)+\lambda=3.5$;
$n(\pi^-\to D^+,\bar D^0)=1-2\alpha_{D^*}(0)-2\alpha_\rho(0)+2+\lambda=4.5$.
The term $\lambda\approx 2\alpha'\la|t|\ra=0.5$ is a result of the $t$-integration. The intercept and slope of the $D^*$ trajectory was evaluated in \cite{ebert} at $\alpha_{D^*}(0)= -1$,  and $\alpha^\prime_{D^*}=0.5\GeV^{-2}$.
So, for the reactions measured in \cite{selex} with the pion beam we chose $n_\pi=4$. Similarly we estimated $n_p=5.5$ and $n_\Sigma=6.5$.
Then, we are in a position to add up the two mechanisms of charmed hadron production in the effective atomic number
\beq
\tilde A_{eff}(x_F)=(1-\delta)\,A_{eff}(x_F)+\delta\,x_F^m\,A,
\label{190}
\eeq
where $A_{eff}(x_F)$ for the conventional mechanism of light quark fragmentation was calculated earlier (dashed curves in Fig.~\ref{alpha1}). All the parameters have been evaluated above, except $\delta$ which we treat as a fitting parameter. However, the experimental errors at large $x_F$ are too big for a statistical fit. One can get a fair description of the data within the range of $\delta= 0.5-0.9$. The solid curves depicted in Fig.~\ref{alpha1} correspond to $\delta=0.8$.

As far as $\delta$ is known, one can try to estimate the IC probability $P_{IC}$. The total cross section of charm production from IC can be presented as $P_{IC}\sigma_{IC}$, where $\sigma_{IC}$ is a part of the total inelastic cross section in which the IC component is resolved and freed. Comparing this with the above estimates one arrives at
\beq
P_{IC}\sim \frac{\delta}{1-\delta}\ B\left({3\over5}n,n\right)\,\frac{\sigma_0}{\sigma_{IC}}.
\label{200}
\eeq
Here $B(x,y)$ is the Beta Function; $\sigma_0$ is the factor in the $x_F$-dependent cross section of inclusive production of charmed hadrons,
$d\sigma(hp\to h_cX)/dx_F=\sigma_0(1-x_F)^n$. We use the data from the LEBC-EHS experiment \cite{data-xF}
for $D^-,\ D^0$-meson production in $pp$ collisions at $400\GeV$ at CERN SPS, which were fitted in \cite{data-xF} with $n=5.4\pm1.2$, in good agreement with our evaluation $n_p=5.5$. Summing up these two channels and doubling the result to account for $\bar D$ production, we get $\sigma_0=170\,\mu b$. Production of $\Lambda_c$ from hadronization of a $c$-quark is strongly suppressed at $x_F\to1$ and can be neglected.

 The magnitude of $\sigma_{IC}$ dependents on a model for the structure of IC. The smallest cross section,
 i.e., the largest estimate Eq.~(\ref{200}), would correspond to an IC of perturbative origin with a $\bar cc$ separation of the order of $1/m_c$. In this case $\sigma_{IC}\sim 1\mb$, and  the range of possible IC weights would be  $P_{IC}\sim 0.0009-0.008$. If the IC had a nonperturbative origin, and the $\bar cc$ were sitting in a nonperturbative potential, the typical cross section $\sigma_{IC}$ would be like for $J/\Psi$ \cite{yuri}, $\sigma_0\sim 5mb$. Correspondingly, $P_{IC}\sim 0.0002-0.002$.
 This estimate is much smaller than the previous evaluation in Ref.~\cite{vogt}
 based on the EMC measurements of charm distribution in the proton.

{\it Charmonium production.}
The penetrating IC component may also show up in the charmonium production as an enhanced $A$-dependence at large $x_F$. However, in this case IC cannot dominate the endpoint region of $x_F\to1$ where the valence quark contribution takes over. Therefore an anomalous nuclear dependence, if any, should be a weaker effect.
Fig.~\ref{vertex}b illustrates the competing mechanisms of charmonium production in terms of the Regge approach.
The first vertex corresponds to $q\bar q$ annihilation, and the related Regge intercept is $\alpha_{R_{3q}}(0)=\alpha_N(0)=-1/2$.

In the second vertex in Fig.~ \ref{vertex}b, all three valence quark of the proton (the system $X_{3q}$) get a low fractional momentum, while the main fraction of the proton momentum is carried by the $\bar cc$. It does not make any difference whether this pair
was created perturbatively, $g\to\bar cc$, or preexisted in the proton as IC. The $t$-channel exchanges and related $x_F$ dependences at $x_F\to1$ are identical, like in the case of open charm production (see above).
The color-octet $\bar cc$ pair converts to $J/\Psi$ via gluonic exchange $X_g$. Thus, instead of a color triplet-antitriplet system, in this case a color octet-octet dipole, $\{X_{3q}\}_8-\{X_g\}_8$, is exchanged. In $1/N_c$ approximation this octet dipole can be presented as two triplet dipoles ${2q}-q$ and $q-\bar q$. This corresponds to the $\rho$-$N$ Regge cut with the intercept $\alpha_{\tilde R_{3q}}(0)=\alpha_N(0)+\alpha_\rho(0)-1=-1$.
\begin{figure}[htb]
 \includegraphics[width=6cm]{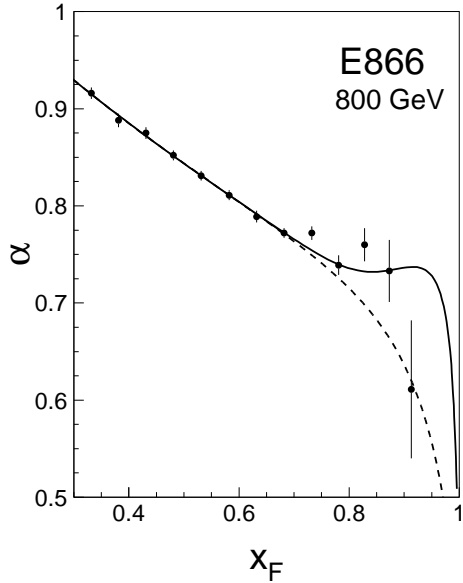}
\caption{\label{e866} The exponent characterizing the $A^\alpha$-dependence of the cross section of $J/\Psi$ production as function of $x_F$. The dashed curves show the result of the fit to the E866 data \cite{e866} by Eq.~(\ref{250}) and extrapolation to larger $x_F$.
The solid curve corresponds to inclusion of IC.}
 \end{figure}

The endpoint $x_F$ dependence of $J/\Psi$ production,
$d\sigma/dx_F\propto (1-x_F)^{n_\psi}$, correlates with the
mechanism: $n_\psi^{q}=1-2\alpha_N(0)+\lambda_\psi=4$; $n_\psi^{\bar
cc}=3-2\alpha_N(0)-2\alpha_\rho(0)+\lambda_\psi=5$. Here
$\lambda_\psi=2\alpha^\prime_N\,\la|t|\ra_\psi=2$ for
$\la|t|\ra_\psi=1.1\GeV^2$ \cite{e789}. Thus,
$n_\psi^{q}<n_\psi^{\bar cc}$, so the $\bar qq$ annihilation, rather
than IC, dominates at $x_F\to1$. However, at medium high $x_F$ data
favor the $(1-x_F)^5$ dependence \cite{na3} indicating the dominance
of the $\bar cc$  mechanism. The latter includes production of a
$\bar cc$ either by projectile gluons, or from the IC. Both
contributions have the same end-point behavior at $x_F\to1$, but the
IC part declines at smaller $x_F$, since it peaks at $\la
x_{IC}\ra=0.75$. To comply with this restrictions we choose the
following form of the IC part of the $J/\Psi$ production cross
section, \beq \frac{d\sigma_{IC}}{dx_F} \propto x_F^{3n^{\bar
cc}_\psi}(1-x_F)^{n^{\bar cc}_\psi}. \label{240} \eeq

The mechanism of nuclear suppression of $J/\Psi$, which is still under debate,
is not the main focus of this paper. Therefore we simply fit the data by the gluon fusion and quark annihilation mechanisms
at $0.3<x_F<0.7$, where the IC contribution is expected to be small,
\beq
A^{J/\Psi}_{eff}(x_F)\,\propto\,
\frac{\epsilon}{1-x_F}\,A_{eff}^q(x_F) +
(1-\delta)\,A_{eff}^{\bar cc}(x_F),
\label{250}
\eeq
with fixed parameters $\delta=0.8$ and $\epsilon=0.04$ (fitted to the absolute cross section \cite{e789}).
$A_{eff}^q(x_F)$ was calculated in analogy to what we have done for open charm (without IC, and with added $c$-quark shadowing calculated in \cite{kth}). $A_{eff}^{\bar cc}(x_F)$ was parametrized as $\ln(A_{eff}^{\bar cc}/A)=a+bx_F+cx_F^2$. The result of the fit with $a=1.11$, $b=-0.56$, $c=0.27$ is plotted by dashed curve in Fig.~\ref{e866}.

Now we can add into Eq.~(\ref{250}) a third term presenting the IC contribution, which according to Eq.~(\ref{240}) has the form $\delta\,x_F^{3n^{\bar cc}_\psi}A$. The final result plotted by solid curve in Fig.~\ref{e866} does not contradict the E866 data, although it is difficult to say with certainty that the data need it.
Notice that the E789 experiment on $J/\Psi$ production \cite{e789} also observed some disorder in $A$-dependence at large $x_F$, which might be a result of the IC contribution.

{\it Summarizing,} the anomalous nuclear dependence for charmed hadrons produced at large $x_F$ gets a natural interpretation in terms of the projectile intrinsic charm which does not attenuate in a nuclear medium. Calculations performed here are in a good accord with data, provided that the probability of IC in the hadronic wave function ranges from $0.1\%$ to $1\%$. A similar contribution of IC to $J/\Psi$ production also causes an enhanced $A$-dependence at large $x_F$, which also agrees with data.

\begin{acknowledgments}

We are thankful to Stan Brodsky who inspired us for this study. We
are also grateful to J\"urgen Engelfried who provided us with
numerous details of the SELEX data. This work was supported in part
by Fondecyt (Chile) grants 1090236, 1090291 and 1100287, and by DFG
(Germany) grant PI182/3-1.

\end{acknowledgments}


\begin{thebibliography}{99}

\bibitem{selex}
  A.~Blanco-Covarrubias {\it et al.}  [SELEX Collaboration],
  Eur.\ Phys.\ J.\  C {\bf 64}, 637 (2009).

\bibitem{forward}
  B.~Z.~Kopeliovich, J.~Nemchik, I.~K.~Potashnikova, M.~B.~Johnson and I.~Schmidt,
  Phys.\ Rev.\  C {\bf 72}, 054606 (2005)

\bibitem{mueller}
  A.~H.~Mueller,
  Phys.\ Rev.\  D {\bf 2}, 2963 (1970).

\bibitem{kancheli}
  O.~V.~Kancheli,
  Pisma Zh.\ Eksp.\ Teor.\ Fiz.\  {\bf 11}, 397 (1970).

\bibitem{agk}
  V.~A.~Abramovsky, V.~N.~Gribov and O.~V.~Kancheli,
  Yad.\ Fiz.\  {\bf 18} (1973) 595
  [Sov.\ J.\ Nucl.\ Phys.\  {\bf 18} (1974) 308].

\bibitem{broad}
  B.~Z.~Kopeliovich, I.~K.~Potashnikova and I.~Schmidt,
  arXiv:1001.4281 [hep-ph], to appear in Phys.\ Rev.\  C.

\bibitem{mv}
  L.~D.~McLerran and R.~Venugopalan,
  Phys.\ Rev.\  D {\bf 49}, 2233 (1994);
  Phys.\ Rev.\  D {\bf 49}, 3352 (1994);
  Phys.\ Rev.\  D {\bf 50}, 2225 (1994).

\bibitem{brodsky} R.~Blankenbecler and S.J.~Brodsky, Phys.Rev. D {\bf 10}, 2973
(1974).

\bibitem{gribov} V.~N.~Gribov,
  Sov.\ Phys.\ JETP {\bf 29}, 483 (1969)
  [Zh.\ Eksp.\ Teor.\ Fiz.\  {\bf 56}, 892 (1969)].

\bibitem{mine}
  B.~Z.~Kopeliovich,
  Phys.\ Rev.\  C {\bf 68}, 044906 (2003).

\bibitem{kps1}
  B.~Z.~Kopeliovich, I.~K.~Potashnikova and I.~Schmidt,
  Phys.\ Rev.\  C {\bf 73}, 034901 (2006).
  
\bibitem{selex-tot}
  U.~Dersch {\it et al.}  [SELEX Collaboration],
  Nucl.\ Phys.\  B {\bf 579}, 277 (2000)

\bibitem{ic1}
S.~J.~Brodsky, C.~Peterson and N.~Sakai,
Phys.\ Rev.\ D {\bf 23}, 2745 (1981).

\bibitem{ic2}
S.~J.~Brodsky, P.~Hoyer, C.~Peterson and N.~Sakai,
Phys.\ Lett.\ B {\bf 93}, 451 (1980).

\bibitem{ic3}
J.~Pumplin,
arXiv:hep-ph/0508184.

\bibitem{ic4}
  J.~Pumplin, H.~L.~Lai and W.~K.~Tung,
  Phys.\ Rev.\  D {\bf 75}, 054029 (2007).
 
 \bibitem{dokshitz}
  Y.~L.~Dokshitzer and D.~E.~Kharzeev,
  Phys.\ Lett.\  B {\bf 519}, 199 (2001).
  
  \bibitem{higgs}
  S.~J.~Brodsky, B.~Kopeliovich, I.~Schmidt and J.~Soffer,
  Phys.\ Rev.\  D {\bf 73}, 113005 (2006);
S.~J.~Brodsky, A.~S.~Goldhaber, B.~Z.~Kopeliovich and I.~Schmidt,
  Nucl.\ Phys.\  B {\bf 807}, 334 (2009).
    
 \bibitem{ebert}
  D.~Ebert, R.~N.~Faustov and V.~O.~Galkin,
  arXiv:0910.5612 [hep-ph].

\bibitem{data-xF}
  M.~Aguilar-Benitez {\it et al.}  [LEBC-EHS Collaboration],
  Z.\ Phys.\  C {\bf 40}, 321 (1988).

\bibitem{yuri}
  J.~Hufner, Yu.~P.~Ivanov, B.~Z.~Kopeliovich and A.~V.~Tarasov,
  Phys.\ Rev.\  D {\bf 62}, 094022 (2000)

\bibitem{vogt}
  B.~W.~Harris, J.~Smith and R.~Vogt,
  Nucl.\ Phys.\  B {\bf 461}, 181 (1996).

\bibitem{e866}
  M.~J.~Leitch {\it et al.}  [FNAL E866/NuSea collaboration],
  Phys.\ Rev.\ Lett.\  {\bf 84}, 3256 (2000).
  
\bibitem{e789}
  M.~S.~Kowitt {\it et al.},
  Phys.\ Rev.\ Lett.\  {\bf 72}, 1318 (1994).

\bibitem{na3}
  J.~Badier {\it et al.}  [NA3 Collaboration],
  Z.\ Phys.\  C {\bf 20}, 101 (1983).

\bibitem{kth}
  B.~Kopeliovich, A.~Tarasov and J.~H\'ufner,
  Nucl.\ Phys.\  A {\bf 696}, 669 (2001).


\end{thebibliography}
\end{document}